\documentclass[12pt,epsfig]{article}
\usepackage{graphicx}

\begin{document}

\baselineskip=18.6pt plus 0.2pt minus 0.1pt

 \def\be{\begin{equation}}
  \def\ee{\end{equation}}
  \def\bea{\begin{eqnarray}}
  \def\eea{\end{eqnarray}}
  \def\nn{\nonumber\\ }
\newcommand{\nc}{\newcommand}
\nc{\bib}{\bibitem} \nc{\cp}{\C{\bf P}} \nc{\la}{\lambda}
\nc{\C}{\mbox{\hspace{1.24mm}\rule{0.2mm}{2.5mm}\hspace{-2.7mm} C}}
\nc{\R}{\mbox{\hspace{.04mm}\rule{0.2mm}{2.8mm}\hspace{-1.5mm} R}}

\begin{titlepage}
\title{
\begin{flushright}
 {\normalsize \small
GNPHE/0902 }
 \\[1cm]
 \mbox{}
\end{flushright}
{\bf On Local  F-theory  Geometries  and  }\\[.3cm]
{\bf  Intersecting D7-branes }
\author{Rachid Ahl Laamara$^{1,4}$\thanks{\tt{doctorants.lphe@fsr.ac.ma}},
Adil Belhaj$^{2,4}$\thanks{\tt{belhaj@unizar.es}}, Luis J.
Boya$^3$\thanks{\tt{luisjo@unizar.es}},
Antonio  Segui$^3$\thanks{\tt{segui@unizar.es}}\\[.3cm]
{\small $^1$ Lab/UFR-Physique des Hautes Energies, Facult{\'e} des Sciences, Rabat, Morocco}\\
{ \small $^2$ Centre National de l'Energie, des Sciences et des Techniques Nucl\'eaires,  CNESTEN} \\
{\small  Cellule Science de la Mati\`ere, Rabat, Morocco} \\{ \small
$^3$ Departamento de F\'{\i}sica Te\'orica, Universidad de Zaragoza,
E-50009-Zaragoza, Spain}
\\
{\small $^4$ Groupement National de Physique des Hautes Energies,
GNPHE}
\\[-6pt] {\small  Si\`{e}ge focal: Facult\'{e} des
Sciences, Rabat,  Morocco }\\
 } } \maketitle
\thispagestyle{empty}
\begin{abstract}
We  discuss  local F-theory  geometries  and theirs gauge theory
dualities in terms of intersecting
 D7-branes wrapped four-cycles in Type IIB  superstring. The
 manifolds are built
as  elliptic K3 surface  fibrations over   intersecting $F_0={\bf
CP^1}\times {\bf CP^1}$ base geometry according to $ADE$ Dynkin
Diagrams. The base  is obtained by blowing
 up   the extended  $ADE$ hyper-K\"{a}hler singularities   of eight dimensional manifolds considered as
   sigma model target spaces  with eight supercharges. The resulting
  gauge theory  of such  local F-theory  models are  given  in  terms of   Type IIB D7-branes
   wrapped intersecting $F_0$. The  four
   dimensional $N=1$ anomaly cancelation
requirement  translates into a condition on the associated affine
Lie algebras.
\end{abstract}
{\tt  KEYWORDS}: F-theory, Type II superstrings, $ADE$
singularities, Supersymmetric Sigma Models.
\end{titlepage}
\newpage

\section{Introduction}

A very nice  way to get supersymmetric   gauge theory from
superstrings, M or F-theory, is to  use  the so-called geometric
engineering method  which is based on
   singular manifold compactifications \cite{KKV}-\cite{Vafa03}. In this way,
 the internal manifold is a  K3 fibration over a base space $B$  which depends  on the theory
 in question. The gauge group $G$   comes from  the singularities of the fiber  while the matter
  is obtained from  non-trivial geometries   in the  base  \cite{KV}.   In the  case of Type
   II superstrings with eight supercharges,  the
complete set of physical parameters of the corresponding   quantum field theory (QFT) is related to the
geometric moduli space of the internal manifold.  The latter is  realized as
a  singular K3 fibration  over a   $\bf CP^1$
complex curve or a collection of intersecting $\bf CP^1$ curves according to Dynkin  geometries.
The corresponding  four-dimensional ($4D$) $N=2$ QFT are represented by quiver  graphs  similar
to Dynkin diagrams of  ordinary, affine,  and indefinite Lie algebras \cite{ KMV,BFS,AABDS,ABS1,ABS2}.

More recently four dimensional  gauge theories with  only four
supercharges, obtained from  local F-theory  models,  have attracted
a lot of attention,
 since they have  connections with the physics of standard model,  grand unification theory
 (GUT) ,
 and Large  Hadron Collider
 (LHC)\cite{Vafa1}-\cite{BO}. For
 instance,
 in this study of   F-theory GUT models involves
  geometric singularities which are  related to D7-branes of  Type IIB  superstring  and are  localized in
   two dimensional  transverse space.  It has been shown that the gauge fields of the MSSM  are  obtained
    from the eight-dimensional worldvolume of a seven-brane wrapping  del Pezzo surfaces  with a
     GUT gauge group. The latter can be    broken to
$SU(3)_C \times SU(2)_L \times U(1)_Y$ via an internal flux through
the seven-brane in the $U(1)_Y$ direction of the GUT. Alternative
study  has been also  done for  M-theory on seven dimensional
manifolds admitting  $G_2$ metrics \cite{AB,ABKKS,AW}.

 \par
  The
aim of this   work  is to contribute  to  F-theory compactification
activities. In particular, we
  engineer  geometrically  quiver   gauge theories   with bi-fundamental matters from local  F-theory    models.
   First, we construct such local models  as K3 fibrations over  a four-dimensional base space.
    The  base geometry  is obtained by  blowing up  the  extended  $ADE$ hyper-K\"{a}hler singularities
      of eight dimensional manifolds considered as   sigma model target spaces  with eight supercharges.
       Actually, this extends the blow up of the  ordinary $ADE$ K\"{a}hler singularities of local K3 surfaces
       described by sigma models with only four supercharges.  Up some details,
       the base in local F-theory  geometry  has been identified with intersecting  $F_0={\bf CP^1}\times
       {\bf CP^1}$ according to the
        $ADE$ Dynkin  diagrams. This means that instead of having  intersecting  ${\bf CP^1}$  curves
         as in  the  case of  $N=2$ geometric engineering, we have now  intersecting  $F_0$ in the base
          of local  F-theory  Calabi-Yau fourfolds. The corresponding    gauge theories   have  been discussed
          in terms of    intersecting  D7-branes in Type IIB superstring. This  allows us  to give   a
           product of $SU$ gauge  groups with bi-fundamental matter encoded in   affine $ADE$
           Dynkin geometries required    by  4D anomaly cancelation condition.  In this way,
           the rank of each $SU$ gauge group is related  to  the Coxter  number on
the corresponding node. In the end of this work, a  discussion on
the standard model gauge group  is given in terms of  twisted non
simply laced $\hat {G}_2^2$  base geometry.

The organization of this paper  is as follows. In section 2, we
engineer geometrically local F-theory models  with $ADE$ base
geometries.  Using $N=4$ sigma model, the manifolds are constructed
as elliptic K3 surfaces fibred over intersecting
  $F_0={\bf CP^1}\times {\bf CP^1}$ according to the   $ADE$ Dynkin  diagrams.
   In section 3,  we  discuss the  corresponding gauge group and matter
content in terms of intersecting D7-branes in Type IIB superstring.
In particular we show how the anomaly cancelation condition
translates into a condition on the associated affine  $ADE$
 geometries in the base of our local F-theory  geometry. A speculation on the standard
 model gauge symmetry
  is given in section 4.  In  particular,   the gauge group  $G =U(1) \times SU(2) \times SU(3)$
   is  discussed in terms of  the  twisted non simply laced  $\hat {G}_2^2$ quiver
   diagram.

\section{On  Constructing  Local F-theory   Models}
In this section, we engineer  local F-theory  models   leading  to
bi-fundamental   matter in four dimensions with $N=1$ supersymmetry.
This will be  given by  an   elliptic K3 fibration over   a
four-dimensional base space $B_4$ with
 the following  Hodge number  condition
\begin{equation}
h^{2,0}(B_4)=0.
\end{equation}
As in  Type IIA superstring and M-theory case  mentioned in the
introduction, the incorporation of the matter may be achieved by
introducing a non-trivial geometry in the base space $B_4$. This
leads us to consider a local model with intersecting base geometry
in order to produce a product gauge group with bi-fundamental
matter. \\ Our strategy in this  section consists of two steps.
First we recall some aspects of F-theory. Then we turn to the
building of local F-theory geometries. Particular emphasis will be
put on the four-dimensional $ADE$ base geometry. The later  is
embedded in a real  eight dimensional manifolds  admitting
hyper-K\"{a}hler metrics. The $ADE $  part will correspond to a
4-dimensional geometry obtained by resolving $  ADE$ extended
hyper-K\"{a}hler singularities studied in \cite{BS}.
\subsection{ Generalities on F-theory}
F-theory defines a non-perturbative vacuum of Type IIB superstring
theory in which the dilaton and axion fields of the superstring
theory are considered dynamical \cite{VafaF,BoyaF}. This introduces
an extra complex modulus which is interpreted as the complex
parameter  of an elliptic curve thereby
 introducing a non-perturbative vacuum of the Type IIB superstring in a twelve-dimensional space-time.
Following Vafa \cite{VafaF}, one may interpret the complex scalar
field  $\tau$  of Type IIB superstring
  as the complex structure of
an extra torus $T^2$  resulting in the aforementioned
twelve-dimensional model. From this point of view, Type IIB
superstring theory may be seen as the compactification of F-theory
on $T^2$. Starting from F-theory, one can similarly look for new
superstring models in lower dimensions obtained by compactifications
on elliptically fibered Calabi-Yau manifolds. For example, the
eight-dimensional F-theory on elliptically fibred K3 is obtained by
taking a two-dimensional complex compact manifold given by \be y^2 =
x^3 + f(z)x + g(z) \ee where $f$  and $g$  are polynomials of degree
8 and 12, respectively. One varies the $\tau$ over the points of a
compact space which is taken to be a Riemann sphere $\bf CP^1$
parameterized by the local coordinate $z$. In other words, the
two-torus complex structure $\tau(z)$  is now a function of $z$  as
it varies over the $\bf CP^1$  base of the above K3 surface. The
above compact manifold generically has 24
 singular points corresponding to $\tau(z) \to \infty$. These singularities have a remarkable physical
  interpretation as each one of the 24 points is associated with the location of a D7-brane in
non-perturbative Type IIB superstring theory. \par  In the
following, we will consider F-theory on elliptic K3 fibration over
$B_4$. In particular, we will focus our attention on the base $B_4$
which gives information about the gauge group form in 4D. This gauge
theory shares similar features as the quiver gauge models  which
describe D-branes at singularities in Type II superstrings on local
Calabi-Yau threefolds.

\subsection{ Base Geometry from  $  ADE$    Extended Hyper-K\"{a}hler  Singularities}
Our choice  of  the  base $B_4$  is motivated  by the  construction
of intersecting  base  geometries
 involved in the geometric engineering  of Type IIA local  models   in  $4D$  with $N=2$ \cite{KKV,KMV,KV}.
  This
 has been obtained  by blowing up  the ordinary $ADE$  singularties of the  K3 surface  which  have a
  nice  physical representation in terms  of sigma model with four   supercharges\cite{W,ABDS,BDS}. However,
    the  idea of the present  construction is to consider   intersecting   geometries  in the base  $B_4$
   by resolving  of the so-called extended  $ADE$  hyper-K\"{a}hler singularities  of
     eight dimensional manifolds  described by  $N=4$ sigma  models \cite{BS}.  To do so, let   us
      first  recall the ordinary  $ADE$  singularities in the case of $N=2$ sigma model. Indeed, consider
       the leading example: $A_1$ singularity. This  has a nice physical representation
       in terms of  two-dimensional  $N = 2$ linear sigma model with   only  $ U(1)$  gauge group  and
three chiral  fields $\phi_i,\; i=1,2,3$ with charges   $q_i = (1,-2, 1)$ satisfying  the local Calabi-Yau condition
\begin{equation}
\sum_i q_i = 1 - 2 + 1 = 0.
\end{equation}
The  $U(1)$ gauge  invaraint
 $x=\phi_1^2\phi_2$,
 $y=\phi_3^2\phi_2$ and
$ z=\phi_1\phi_2 \phi_3$
satisfies the usual the  $A_1$ singularity
\be
xy=z^2.
\ee
The corresponding $D$-term bosonic potential $U(\phi_1, \phi_2,\phi_3)$,
  in supersymmetric theories with four supercharges, reads as
\begin{equation}
\label{sigma12}
 U(\phi_1, \phi_2,\phi_3)=(|\phi_1|^2+|\phi_3|^2-2|\phi_2|^2-R)^2.
\label{U}
\end{equation}
In this equation, $R$ is the coupling parameter of the  $U(1)$ Fayet-Iliopoulos (FI)
 term one may introduce in the Lagrangian model. In the superfield language, the action then reads
\be
 S(\Phi,V)= \int d^2xd^4\theta \bar\Phi e^V \Phi -R\int d^2xd^4\theta V,
\label{S} \ee where $\Phi$ and $V$ are respectively the chiral and
gauge superfields. The presence  of the FI term resolves  the
singularity of the potential $U(\phi_1, \phi_2,\phi_3)$.
Geometrically, this corresponds to replacing the singular point
$x=y=z=0$ by a $\bf CP^1$  described  by
\begin{equation}
\label{cp1}
|\phi_1|^2+|\phi_3|^2=R.
\end{equation}
  Note that $\phi_2$
defines the cotangent direction  over $\bf CP^1$   of the deformed  $A_1$
 ALE space\footnote{More general   study for  $ADE$  singularities of K3 surface can be found in the  appendix.}.

The above  $N=2$ sigma model can be  extended  to $N = 4$
supersymmetric gauge theory with eight  supercharges describing
eight dimensional manifolds with $ADE$ hyper-K\"{a}hler
singularities. This involves $ U(1)^r$ gauge symmetry and
 $r+2$ hypermultiplets with a  matrix charge $Q^a_i$, which  can be identified
  with the  Cartan matrices of $ADE$  Lie algebras \cite{BS}. The $N = 4$ D-flatness
   equations of such models are generally given by the  moduli space of the hypermultiplets vacua
\be \label{sigma4} \sum_{i=1}^{r+2}Q_i^r[\phi_i^\alpha
{\bar{\phi}}_{i\beta}+\phi_i^\beta {\bar{\phi}}_{i\alpha}
]=\vec{\xi}_a \vec{\sigma}^\alpha_\beta, \ee where
$\phi_i^{\alpha}$'s
          denote   $r+2$   component field doublets of
          hypermultiplets, $\vec \xi_a$ are  $r$ FI 3-vector
 couplings  rotated by $SU(2)$
          symmetry, and
           $\vec \sigma^\alpha_\beta$ are the
          traceless $2\times 2$ Pauli matrices. Equations  (\ref{sigma4})
    deal with the hypermultiplet branch and give a gauge invariant
hyper-K\"{a}hler target space. For each $U(1)$ factor of the
$U(1)^r$ gauge group, they
 involve a triplet of FI parameters.  Note also that
(\ref{sigma4}) have a manifest $SU(2)_R$ symmetry which is absent in
$N=2$ sigma model discussed before.\\ Using the $SU(2)_R$
transformations
  \be \phi^{\alpha}=\varepsilon^{\alpha\beta}\phi_
{\beta},\quad\overline{(\phi^\alpha)}=\overline{\phi}_\alpha,\quad
  \varepsilon_{12}=\varepsilon^{21}=1,
   \ee
 and  replacing the Pauli matrices by their expressions, the identities
  (\ref{sigma4}) can be
   split as
  follows  \bea
\label{extendedsigma}
 \sum\limits_{i=1}^{r+2} Q_i^a(
  |\phi^1_i|^2-|\phi^2_i|^2) &= &\xi^3 _a\nn \sum\limits_{i=1}^{r+2}Q_i^a
  \phi^1_i \overline{\phi}_{i}^2&=&\xi^1_a+i{\xi^2}_a
  \\
  \sum\limits_{i=1}^{r+2} Q_i^a\phi^2_i
  \overline{\phi}_{i}^1&=&\xi^1_a-i{\xi^2}_a. \nonumber \eea
We will see later that the  solution of these equations defines
cotangent  bundles over
  intersecting 4-cycles. Actually, this  extends  the result of the ordinary $ADE$ singularities
   of the K3 surface where it  appears  intersecting 2-cycles. These 4-cycles will be  identified
with  the  base of local F-theory  models  allowing  us  to produce
a product of $SU$ gauge groups  with bi-fundamental  matter. The
latter can be encoded in   quiver diagrams similar to Dynkin graphs.
For simplicity reasons we center on  the $A_r$ geometry. In this
case, we have   a  matrix charge  of  the form
\begin{equation}
Q^a_i=\delta^a_{i-1 }-2 \delta^a_i+ \delta^a_{i+1}.
\end{equation}
To handle the   corresponding  D-terms equations, it should be
interesting to note that they are quite similar to the analysis of
$N=2$  sigma model discussed in the appendix.  Indeed, forget for
the while  the two last equations of (\ref{extendedsigma})  for
$A_r$ model. Setting  $\xi^3_a = R_a-P_a$   and taking the condition
that $R_a> P_a$, one may put
  the  first  equation of  (\ref{extendedsigma}) for $A_r$ model  into the two following equations
\bea \label{extendedA2}
 |\phi^1_{ a-1}|^2 + |\phi^1_{a+1}|^2 -2|\phi^1_a|^2=R_a\nn
|\phi^2_{a-1}|^2 + |\phi^2_{a+1}|^2 -2|\phi^2_a|^2 = P_a.
\eea
 Let us comment these equations:
\begin{itemize}
\item
They describe two orthogonal copies of $A_r$ models.  In the case
where $R_a = 0$ and $P_a$ positive definite, or $R_a$ positive
definite and  $P_a = 0$, one of the  $A_r$  models  becomes
singular. If   $R_a = P_a = 0$  both of them are singular.
\item For $R_a$ positive definite and  $P_a$
positive definite, we  have  the blown up of the two $A_r$ singularities  where
each  $A_r$ model has four charges   obtained  by breaking  $N= 4$ supersymmetry to $N=2$.
\end{itemize}
To better understand the geometry given by (\ref{extendedsigma}) for $A_r$ model, let us deal with the leading
example corresponding to $r=1$.
 This involves one   $U(1)$  gauge group   with three hypermultiplets  $\phi_i$ of charges  $ Q_i=(1,-2,1)$.
  In this case, (\ref{extendedA2})  reduce to
\bea \label{An=4}
 |\phi^1_1|^2 + |\phi^1_{3}|^2 - 2|\phi^1_2|^2=R \nn
|\phi^2_{1}|^2 + |\phi^2_{3}|^2 - 2|\phi^2_2|^2 = P. \eea  These
equations  describe the  cotangent bundle over the  Hirzebruch
surface $F_0={\bf CP^1}\times {\bf CP^1}$. Indeed, putting
$\phi^1_2=\phi^2_2=0$, (\ref{An=4}) become  \bea
 |\phi^1_1|^2 + |\phi^1_{3}|^2 =R\nn
|\phi^2_{1}|^2 + |\phi^2_{3}|^2 = P. \eea These two equations define
the Hirzebruch surface  $F_0={\bf CP^1}\times {\bf CP^1}$
generalizing
 (\ref{cp1}).  Now  we  come back to the two last equations of  (\ref{extendedsigma}) and see theirs roles.
 For simplicity reason, let us consider  the case of  $U(1)$  gauge symmetry and  assume that
 the corresponding charges
 $Q_i$ are all positive which can be obtained by possible exchanging some chiral fields in  the generic case.
Putting  $\xi_i^2=\xi_i^3=0$ and  introducing
 $x_i=\sqrt{Q_i}\phi^1_i$  and $y_i=\sqrt{Q_i}\phi^2_i=0$, the  two last  equations (\ref{extendedsigma}) can be rewritten as
 \bea
\label{extendedA} x_i\bar{y_i}&=&0\nn \bar{x_i}y_i &=& 0. \eea They
can  be viewed as a scalar product showing two orthogonal variables.
Based on this remark  and  the fact that  $\phi^1_2$ and $\phi^2_2$
describe two orthogonal non compact directions  over $F_0$, our
total geometry is given  by
   the cotangent   fiber over   $F_0$
   base. Locally,  up to an orbifold action the fiber  can be
   identified with local K3 surfaces making contact with D7-brane physics in Type IIB
superstring. By assumption, the cotangent fiber can be divided by
subgroups of $SU(2)$. If we denote this subgroup by $\Gamma$, the
fiber looks like \be {\bf C}^2/\Gamma \ee leading to $ADE$ gauge
symmetries in eight dimensions \cite{VafaF}.

The story is similar for the general $A_r$ geometry   where we
obtain intersecting $F_0$'s  where  all $\xi^a$'s are no zero. We
thus
   expect to obtain  the cotangent bundle over $r$  intersecting  $F_0={\bf CP^1}\times {\bf CP^1}$
   according to $A_r$ Dynkin diagrams.  This means that the base geometry, of the cotangent bundle,
   consists of $r$ intersecting $F_0={\bf CP^1}\times {\bf CP^1}$   arranged as shown here
\be
    \mbox{
         \begin{picture}(20,30)(70,0)
        \unitlength=2cm
        \thicklines
    \put(0,0.2){\circle{.2}}
     \put(.1,0.2){\line(1,0){.5}}
     \put(.7,0.2){\circle{.2}}
     \put(.8,0.2){\line(1,0){.5}}
     \put(1.4,0.2){\circle{.2}}
     \put(1.6,0.2){$.\ .\ .\ .\ .\ .$}
     \put(2.5,0.2){\circle{.2}}
     \put(2.6,0.2){\line(1,0){.5}}
     \put(3.2,0.2){\circle{.2}}
     \put(-1.2,.15){$A_{r}:$}
  \end{picture}
} \label{ordAk} \ee where the nodes represent $F_0$, while their
intersections are represented by the links. This generalizes the
case of $N=2$ sigma model where     each node  is  associated with a
$ {\bf CP^1}$ of  the $A_r$  deformed singularity.  We will conclude
this construction  by noting that this
     analysis for  $A_r$  model  may be extended to the others $DE$  Lie algebras where the  corresponding geometries
      are classified by Dynkin  graphs.\\
Having constructed the  base  geometry of our local F-theory   model
as intersecting
 $F_0={\bf CP^1}\times {\bf CP^1}$ according to $ADE$  Dynking   graphs, we
will discuss the corresponding quiver  gauge theory in  four
dimensions.

\section{ Intersecting D7-branes and Bi-fundamental Matter}
 Our analysis here will be  based
on a dual Type IIB superstring description in terms of D7-brane
backgrounds. In addition to D7-branes, a F-theory background may
also contain D3-branes  localized at some singular points. However,
for reasons of simplicity, only the world volume of D7-branes will
play a role in models we deal with. Theses branes  wrap 4-cycles and
fill the four-dimensional Minkowski space. Indeed, a local
description of F-theory near the $A_{n-1}$ singularity of the  K3
surfaces is equivalent to $n$ units of D7-branes in Type IIB dual
version in eight
 dimensions \cite{VafaF}.  This  can be obtained by taking $\Gamma$  as
 \be
\Gamma=Z_n.
 \ee
On each D7-brane we have a $U(1)$ symmetry. When the $n$ D7-branes
approach each other, the gauge symmetry is enhanced from
$U(1)^{\otimes n}$ to $SU(n)$ \cite{VafaF}. This can be extended to
arbitrary gauge group involving exceptional D7-branes \cite{Vafa5}.
An extra compactification of F-theory down to four-dimensional
space-time is equivalent, in Type
 IIB superstring side,  to wrapping D7-branes on 4-cycles. Here they  will  can be identified   with intersecting
  ${\bf CP^1}\times {\bf CP^1}$.
Consider now $m$ different stacks of D7-branes. This brane
configuration can be reinterpreted as  singularities of type
$A_{n_i-1}$ which can meet at a point where the singularity jumps to
higher gauge group. Note that each  stack  contains $n_i$ units of
D7-branes. Identifying $m$ with the number of intersecting ${\bf
CP^1}\times {\bf CP^1}$ and using the result of the geometric
engineering  in Type II superstrings and M-theory on $G_2$ manifolds
and assuming  that each stack of D-branes wrap one $F_0$,  the the
gauge group corresponding to such  configurations  can be described
by
\begin{equation}
\label{G}
 G\ =\ \bigotimes_{i=1}^m SU(n_i).
\end{equation}
The  $n_i$  which are  integers  can be specified by  physical
requirements.   It turns out that  in the case of
   $N=1$ models with four supercharges,  these  integers  could  be fixed  by the anomaly cancelation
    condition \cite{HI,FHHI,U}. They should form  a null vector of  a   matter matrix $I_{ij}$:
\begin{equation}
\label{intersection}
 \sum_{i=1}^mI_{ij}n_i\ =\ 0.
\label{acc}
\end{equation}
As in the  $N=2$  scenario \cite{KKV,KMV,ABS1,ABS2},  the gauge
group and matter depend on the intersection  matrix $I_{ij}$. Recall
that the intersection matrix of $n$ real dimensional sub-lagrangian
manifolds considered as the blowup  of singularities in
$n$-dimensional Calabi-Yau manifolds  is symmetric for $n$ even and
antisymmetric for  $n$ odd. In more general geometries, the
intersection matrix may be written as a linear combination of a
symmetric and antisymmetric  term. From the obvious similarity with
the $N=2$ scenario it is  not  surprising to  to see some analogs
with  $ADE$ diagrams in eight-dimensional  hyper-K\"{a}hler
manifolds that we considered  in the previous section. Indeed,  the
intersection theory assigns the intersection number to complex
surfaces inside of such a manifold. For example, the
self-intersection of the zero section in the cotangent bundle of
$F_0$  is equal to its minus Euler characteristic, i.e. $-4$.
Consider now  a lattice of compact 4-cycles  generated by  $F^i_0$.
Assume that  $F^i_0$ intersects $F^{i+1}_0$ at two
 points. This can be supported by the fact that   each     ${\bf CP^1}$ inside  $F^i_0$ intersects just one
  ${\bf CP^1}$ in the next  $F_0^{i+1}$. In this way,  the intersection numbers
  of the
$F_0$'s  can be  given by
 \bea
F_0^i.F_0^i=-4\nn F_0^i.F_0^{i+1} =2,
 \eea
 with others vanishing.  This means that   $F^i_0$ does not intersect $F^j_0$ if $|j - i| > 1$. Endowed
with this intersection form, the lattice of compact 4-cycles  can be
identified with  the root lattice of the $ADE$  Lie groups, up to a
multiplication  factor. A nice geometric interpretation  for $ADE$
diagrams could be also obtained by looking at symplectic resolutions
of  finite quotients $V/G$ where $G$ is a finite subgroup of $Sp(2)$
generated by symplectic reflections (i.e. elements $g$ with fixed
locus of complex codimension 2). It would be worth investigating
such geometries.

From    this discussion,  we  see that  the information on our
intersection geometry is naturally encoded in the Cartan matrix
$K_{ij}$  of Lie Algebras.
The latters are classified into three  categories\cite{Kac}:\\
1. Finite type $(det\; K> 0 )$. In this case,  there exists a real
positive definite vector $u ( u_i > 0; i = 1, 2, ...)$ such that
$K_{ij}u_j = v_j > 0$. \\2. Affine type,
 $corank(K) = 1, det \;K = 0 $. There exists a unique, up to a
multiplicative factor, positive integer definite vector $n$ $( n_i >
0,  i = 1, 2, \ldots)$ such that $K_{ij}n_j =  0$. \\ 3. Indefinite
type $(det\; K \leq 0 ), corank(K) \neq 1$.  There exists a real
positive definite vector $u (u_i > 0; i = 1, 2, \ldots )$ such that
$K_{ij}u_j =
-v_i < 0.$\\
   From this classification, it follows that the anomaly cancelation condition  is translated into a condition
    on the affine Lie algebra. Indeed,   (\ref{acc}) can  be solved by
\bea
I_{ij} &\equiv& -2K_{ij}\nonumber\\
 n_i&\equiv& s_i n\\
m&\equiv& r+1\nonumber \eea where $K_{ij}$ are  now the Cartan
matrices of $ADE$  affine Lie algebras and  $n$ is an arbitrary
number. While $s_i$
 are the Coxeter labels of the associated
affine Lie algebra of rank $r$. As  we have seen,  the latter   is
related to  the number of intersecting  $F_0={\bf CP^1}\times {\bf
CP^1}$ that we should have in the  base of our  local  F-theory
geometry. The   gauge group finally reads as
\begin{equation}
\label{Gf}
 G\ =\ \bigotimes_{i=1}^{r+1} SU(s_in),
\end{equation}
with bi-fundamental chiral matter transforming in $(s_in,s_jn)$
representations.\\
In the end of this section, we would like to note that it is
possible to consider  the two other Lie algebras by adding non
trivial matters in order to satisfy the anomaly cancelation
condition. We expect to  have a similar analysis  made for conformal
invariance in the Type IIA  geometric engineering method of $N=2$ in
four dimensions studied in \cite{ABS1,ABS2}. It should be
interesting to develop this issue in the future\cite{work}.

\section{Discussions}
 In this work, we have engineered  local F-theory geometries and four dimensional  $N = 1$  quiver gauge  theories
 with   bi-fundamental matters. The  manifolds  have  been  built   as   an elliptic K3 fibration
over
  intersecting $F_0= {\bf CP^1} \times {\bf CP^1}$ surfaces   according   $ADE$  geometries. This  intersecting
   base  geometry  can be obtained from the deformation of the extended  of $ADE$  hyper-K\"{a}hler
   singularities
   of eight dimensional manifolds considered as  target spaces of  $N=4$ sigma model. Our main results may
    be summarized as follows:\\
(1) Using two-dimensional  $N = 4$      field theory with $U(1)^r$
gauge and $r+2$  hypermultiplets with a matrix charge, which can be
identified, up some details, with  $ADE$ Cartan matrices,  we have
constructed the $B_4$ geometry  of local Calabi-Yau fourfolds. In
particular, it has been shown that the   corresponding  target
spaces   are  described by the cotangent bundle over intersecting
$F_0= {\bf CP^1} \times {\bf CP^1}$ according to $ADE$  Dynkin
diagrams. The latter
  has been identified with  the base of our local F-theory geometries.\\
(2) By the help of intersecting  D7-branes in Type IIB superstring,
we have discussed the physics content of F-theory on such local
Calabi-Yau fourfolds.  In particular,  we have engineered  the gauge
group
  and matter content.    The $N=1$   $4D$   anomaly cancelation condition  has been converted  into
  a condition on the associated affine Dynkin  geometries  in the base $B_4$.\\
Using this analysis  the  gauge group of standard model of
electromagnetic, weak, and strong interactions
 could  be  discussed in terms of   the folding  of  the   Dynkin diagram of  affine $E_6$ Lie algebra given
  by the following diagram
   \vspace{-3cm}
$$
    \mbox{
         \begin{picture}(40,350)(50,50)
        \unitlength=2cm
        \thicklines
    \put(0,4){\circle{.2}}
     \put(.1,4){\line(1,0){.5}}
     \put(.7,4){\circle{.2}}
     \put(.8,4){\line(1,0){.5}}
     \put(1.4,4){\circle{.2}}
    \put(1.5,4){\line(1,0){.5}}
    \put(2.1,4){\circle{.2}}
    \put(2.2,4){\line(1,0){.5}}
    \put(2.8,4){\circle{.2}}
   \put(1.4,4.1){\line(0,1){.5}}
   \put(1.4,4.7){\circle{.2}}
   \put(1.4,4.8){\line(0,1){.5}}
   \put(1.4,5.4){\circle{.2}}
  \end{picture}
}$$
 \vspace{-7cm}
  \\
   As well known,    $E_6$ Lie algebra\footnote{We thank J. L.Cortes for pointing
   us the reference \cite{R}.} has  some relations  with  grand unified theory, since it    has been
    considered  as a possible gauge group which, after its breaking, gives rise to   the  gauge symmetry
     of the standard model \cite{R}.  However,  our  idea   is bit different since instead of   taking
     this symmetry as a singularity  in the K3 surface  fiber of F-thory compactification     and   break it
      to the usual standard  model gauge group,  we will consider it in  the base  geometry $B_4$.
       In this way, (\ref{Gf})   becomes
\begin{equation}
 G\ = SU(3n)\times SU(2n)^3\times SU(n)^3 .
\end{equation}
To get  the  gauge symmetry of the standard model, we proceed as follows:
\begin{itemize}
\item use the  folding techniques of  simply laced  Dynkin diagram ($SLDD$)
\item take  the limit  $n=1$.
\end{itemize}
To do so,  recall that the folding method gives rise  non simply
laced Dynkin diagrams ($NSLDD$). In particular,  the corresponding
diagram  are   obtained from the simply laced ones by identifying
the Dynkin nodes which are permuted by a  outer-automorphism group
$\Gamma$.  Formally, we can write this correspondence as  follows
\begin{equation}
SLDD/\Gamma\equiv  NSLDD
\end{equation}
It turns out that,   the  Dynkin diagrams of affine  $E_6$  has
different realizations of the outer-automorphism group $\Gamma$
leading to different non simply laced diagrams.  In the case of
$\Gamma=Z_3$, we get the twisted $G^2_2$  affine Dynkin diagram. The
latter has
 the following Coxter numbers
\be s_i=(1,2,3) \ee  and can represented  by the following Dynkin
diagram
\begin{figure}[tbph]
\par
\begin{center}
\hspace{3cm} \includegraphics[width=3.5cm]{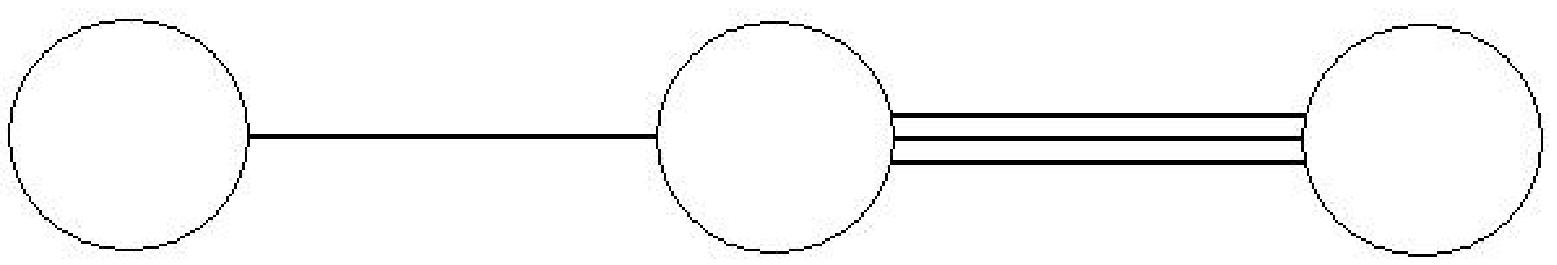}
\end{center}
\end{figure}
\\
    In this case, the (\ref{Gf}) reduces to $
 G\ = SU(3)\times SU(2)\times U(1)$. It should be interesting to come
 back to this observation in the future.

{\bf Acknowledgments.}   RAL would like to thank  A.  Arhrib and  S.
Khalil for discussions on related subjects, scientific helps and
kind hospitality at the BUE Centre for Theoretical Physics.  AB
would like to thank  M. Asorey,  B. Belhorma, J. L. Cortes, I.
Dolgachev, L. B. Drissi, J. McKay,  J. Rasmussen, E. H. Saidi and A.
Sebbar for collaborations,  discussions  on related subjects and
scientific helps. LJB  and AS have been supported by CICYT (grant
FPA-2006-02315) and DGIID-DGA (grant 2007-E24/2). This work has been
supported by Fisica de altas energias: Particulas, Cuerdas y
Cosmologia, A/9335/07.

\section{Apendix}
 The local  K3 surface  with $ADE$ singularities  can  identified with the asymptotically
 locally Euclidean (ALE) space  which is algebraically given by
\begin{equation}
\label{ADE} f_{ADE}(x,y,z) = 0,
\end{equation}
where $(x,y,z)$ are complex variables.   The $ADE$  singularities
are classified by \bea A_{n-1} : f(x, y, z) = xy - z^n\nn D_n : f(x,
y, z) = x^2 + y^2z + z^{n-1} \nn
E_6 : f(x, y, z) = x^2 + y^3 + z^4\\
E_7 : f(x, y, z) = x^2 + y^3 + yz^3\nn E_8 : f(x, y, z) = x^2 + y^3
+ z^5. \nonumber \eea They are all of them  singular at  $x = y = z
= 0$ since it is the only solution of $f_{ADE} = df_{ADE} = 0$.
These geometries can be 'desingularized'  by deforming the complex
structure of the surface or varying its K\"{a}hler structure. This
consists in blowing up the singularity by a collection of
intersecting
 complex curves. This means that  we replace the singular point
$(x,y,z)=(0,0,0)$ by a set of intersecting
 complex curves ${\bf CP^1}$ (two-cycles). The
nature of the set of
 intersecting ${\bf CP^1}$ curves depends  on the type
of the singular surface one is considering. The smoothed  $ADE$
surfaces
 share several features with the  $ADE$ Dynkin diagrams. In particular, the
intersection matrix   of the complex curves used in the resolution
of  the $ADE$  singularities is, up to some details, minus the $ADE$
Cartan matrix $K_{ij}$. As in the case of  $A_1$, the  deformed
$ADE$   K\"{a}hler  geometries  can also have  nice
  physical interpretation   in terms of  $N=2$   sigma models.  In general, these are described
  by a $U(1)^r$ gauge group with   $ (r + 2) $ chiral multiples $\phi_i $ with $q_i^a$  vector charges
   satisfying  the Calabi-Yau condition
\begin{equation}
\sum_i q_i^a = 0,
\end{equation}
 under which the gauge model flow in the infrared to $2D$ $N = 2$ superconformal field theory. The  $ADE$
  spaces of classical vacua, in the absence of the sigma model superpotential, are given by
\begin{equation}
\label{sigma2}
 U(\phi_i)= \sum_i (q_i^a|\phi_i|^2-R^a)^2.
\label{Uade}
\end{equation}
where  $r$ is the rank of  $ADE$  algebras in question. $q_i^a$, up to details, are proportional
 to the corresponding  Cartan matrices  $K_{ai}$.  In the case of  $A_{r}$ ALE space,  we have $ U(1)^{r}$
theory with  $ r+2$ chiral fields, with charges given by \bea q^
1&=& (1,-2, 1, 0, 0, 0, . . . , 0),\nn q^ 2&=& (0, 1,-2, 1, 0, 0, .
. . , 0),\nn
q^ 3&=& (0, 0, 1,-2, 1, 0, . . . , 0),\\
&\ldots&\nn q^{r}&=& (0, 0, 0, 0, . . . , 1,-2, 1).\nonumber \eea
 In this case, the generators of gauge invariant chiral fields are
  $x=\phi_1^{r+1}\phi_2^{r}\phi_3^{r-1}\ldots\phi_{r+2}^{0}$, $y=\phi_1^{0}\phi_2^{1}\phi_3^{2}
  \ldots\phi_{r+2}^{r+1}$ and
  $z=\phi_1\phi_2\phi_3\ldots\phi_{r+2}$. They
satisfy    the   $su(r+1)$ singularity  equation \be xy=z^{r+1} \ee
The deformation of this singularity can be given by  the following
D-terms
\begin{equation}
 |\phi_{a-1}|^2-2 |\phi_a|^2 +|\phi_{a+1}|^2=R^a.
\label{Uade}
\end{equation}

\end{document}